\documentclass[a4paper, 11pt]{article}
\usepackage[T1]{fontenc}
\usepackage{jheppub,graphicx,epsf,amssymb,amsbsy,amsfonts,amssymb,amsmath, empheq,mathrsfs,appendix,soul}
\usepackage{hyperref}

\def\be{\begin{equation}}
\def\ee{\end{equation}}
\def\bea{\begin{eqnarray}}
\def\eea{\end{eqnarray}}

\def\b{\beta}
 
\def\d{\delta}
\def\e{\epsilon}
\def\f{\phi}

\def\s{\sigma}

\def\bg{\bar{g}}
\def\beq{\begin{eqnarray}}\def\eeq{\end{eqnarray}}
\def\ba#1\ea{\begin{align}#1\end{align}}
\def\bg#1\eg{\begin{gather}#1\end{gather}}
\def\bm#1\em{\begin{multline}#1\end{multline}}
\def\bmd#1\emd{\begin{multlined}#1\end{multlined}}

\def\b{\beta}

\def\d{\delta}
\def\D{\Delta}
\def\e{\epsilon}

\def\G{\Gamma}

\def\p{\phi}

\def\s{\sigma}

\def\pa{\partial}

\def\({\left(}
\def\){\right)}
\def\[{\left[}
\def\]{\right]}

\def\tn{\textnormal}

\def\b{\beta}

\def\d{\delta}
\def\s{\sigma} 
\def\om{\omega}

\def\p{\partial}
\def\no{\nonumber}
\def\f{\frac}

\def\D{\Delta}

\newcommand{\mfour}{\widetilde{\mathcal{M}}^{\Phi}_{4}(1^{a_{1},\epsilon_{1}}_{\Delta_{1},-},2^{a_{2},\epsilon_{2}}_{\Delta_{2},-},3^{a_{3},\epsilon_{3}}_{\Delta_{3},+},4^{a_{4},\epsilon_{4}}_{\Delta_{4},+})}

\newcommand{\bz}{\bar{z}}

\title{MHV Gluon Scattering in the Massive Scalar Background and Celestial OPE}

\author[1,3]{Shamik Banerjee,}
\author[1,3]{Raju Mandal,}
\author[2,3]{Akavoor Manu} 
\author[4]{and \ Partha Paul}

\affiliation[1]{National Institute of Science Education and Research (NISER), Bhubaneswar 752050, Odisha, India}
\affiliation[2]{Institute of Physics, Sachivalaya Marg, Bhubaneshwar, India-751005}
\affiliation[3]{Homi Bhabha National Institute, Training School Complex, Anushaktinagar, Mumbai 400094, India}
\affiliation[4]{Centre for High Energy Physics, Indian Institute of Science, C.V. Raman Avenue, Bangalore 560012, India}

\emailAdd{banerjeeshamik.phy@gmail.com}
\emailAdd{rajuphys002@gmail.com}
\emailAdd{manu.akavoor@gmail.com} 
\emailAdd{pl.partha13@gmail.com}

\abstract{In this paper we study the tree-level OPE between two positive helicity outgoing gluons in the celestial CFT for the Yang-Mills theory chirally coupled to a massive scalar background. This theory breaks the translation as well as scale invariance. We compute the subleading terms in the OPE expansion and show that they are same as the subleading terms of the OPE expansions in the MHV sector. As a result the amplitudes of this theory also satisfy the set of differential equations obtained previously for MHV amplitudes in pure YM theory. This is not surprising because the symmetries coming from the leading and subleading soft gluon theorems do not change in the presence of a massive scalar background.}

\begin{document}
\maketitle
\flushbottom
\section{Introduction}

Scattering amplitudes are an important set of observables in any quantum field theory in four dimensional asymptotically flat spacetime. Usually these amplitudes are written in momentum eigen-basis in which translation invariance is manifest. However, instead of the momentum eigenstates if one uses the conformal primary or the boost eigen-basis, scattering amplitudes transform as a correlation function of a 2D conformal field theory under the (Lorentz) $ SL(2,C) $ transformations \cite{Pasterski:2016qvg,Pasterski:2017kqt,Banerjee:2018gce, Freidel:2022skz, Cotler:2023qwh}\footnote{The Lorentz group $SL(2,\mathbb C)$ act on the celestial sphere as the group of global conformal transformations.}. The amplitudes written in the conformal primary basis are known as celestial amplitudes. For massless particles they are obtained by Mellin transformations with respect to the energies of the external particles \cite{Pasterski:2017kqt,Banerjee:2018gce}. 

Now the fact that they transform like conformal correlators in 2D allows us to use powerful techniques familiar from the study of the 2D CFT. In particular, momentum space soft theorems can be reinterpreted as Ward identities for various current algebra symmetries on the celestial sphere \cite{Strominger:2014as, Strominger:2014ass,Banerjee:2020zlg, Donnay:2018neh, Pate:2019mfs, Fan:2019emx, Nandan:2019jas, Adamo:2019ipt, Banerjee:2020vnt, He:2014cra, Guevara:2021abz, Strominger:2021lvk, Strominger:2021mtt, Kapec:2015ena, Kapec:2014zla, Campiglia:2015qka, Nande:2017dba, He:2020ifr,Kapec:2014opa,He:2017fsb,Donnay:2020guq,Stieberger:2018onx,Banerjee:2022wht,Himwich:2021dau}. For example, the leading soft theorem for a gluon can be interpreted as the Ward identity for the $SU(N)$ Kac-Moody algebra \cite{Strominger:2014ass, Donnay:2018neh, Pate:2019mfs, Fan:2019emx, Nandan:2019jas, Adamo:2019ipt, He:2014cra, Kapec:2015ena, Kapec:2014zla, Campiglia:2015qka, Nande:2017dba, He:2020ifr}. Similarly, the subleading soft gluon theorem can also be interpreted as the Ward identity of an infinite dimensional current algebra \cite{Banerjee:2020vnt, Guevara:2021abz, Strominger:2021lvk, Strominger:2021mtt}. The symmetry algebras coming from leading and subleading soft gluon theorems contain null states \cite{Banerjee:2020vnt}. Decoupling of these null states gives rise to a set of differential equations for the celestial amplitude \cite{Banerjee:2020vnt}\footnote{Please see \cite{Hu:2021lrx} for the derivation of the differential equations in the pure YM case using BCFW shift.} \footnote{For a similar story in the case of gravitons please see \cite{Banerjee:2020zlg, Banerjee:2021cly, Banerjee:2023zip}.}. Hence the amplitudes are heavily constrained due to the presence of infinite dimensional symmetries coming from soft gluon theorems. One gets these null states in case of gravitons as well as gluons by demanding consistency of the celestial OPE with the soft theorems \cite{Banerjee:2020zlg, Banerjee:2021cly, Banerjee:2020vnt, Banerjee:2023zip}. For more discussions on celestial OPE please see \cite{Fan:2019emx, Pate:2019lpp, Banerjee:2020kaa, Banerjee:2021dlm, Bhardwaj:2022anh, Raclariu:2021zjz, Pasterski:2021rjz, McLoughlin:2022ljp}.


Now for gluons, the Mellin transform of tree level three and four point scattering amplitudes are distributional in nature due to the momentum conserving delta function and the conformal invariance of the tree level Yang-Mills theory. The momentum conservation constraint can be removed by supplying background momentum to the YM theory which breaks translational invariance explicitly \cite{Fan:2022vbz}.  In \cite{Fan:2022vbz}, this was achieved by coupling the YM theory to background massless dilaton field. But this does not break the conformal invariance of the YM theory. Translation invariance breaking solution was also considered in \cite{Costello:2022wso}.

A somewhat different set up was considered in \cite{Casali:2022fro}. They chirally coupled the Yang-Mills theory to a massive dilaton background. As a result the space-time translation as well as the scale invariance of the tree-level Yang-Mills theory was explicitly broken but the (Lorentz) 2D conformal invariance was preserved. The interesting fact about this coupled theory is that, when written in the celestial basis, the 3-point amplitudes take the usual 2D CFT form. It was shown \cite{Casali:2022fro} that the leading soft gluon theorem and the leading OPE structure remain unchanged for this chirally coupled theory. In this paper we show that the subleading soft gluon theorem and subleading OPE structrures also remain unchanged, giving rise to the same null state relation as obtained in \cite{Banerjee:2020vnt}. We expect this to be true as the leading and subleading soft gluon theorems don't require space-time translation or scale invariance.

This paper is organized as follows. In section \ref{cel_amp}, following \cite{Casali:2022fro} we briefly review the Yang-Mills amplitude chirally coupled to a massive dilaton background and Mellin transform it to get the celestial amplitude. OPE factorisations of the 4-point celestial amplitude is given in section \ref{4point}. In section \ref{sub_th} we show that the subleading soft gluon theorem remains the same for this coupled theory. By demanding the consistency of the OPE at order 1 with the soft theorem we get the same null state relation under the soft current algebra as obtained in \cite{Banerjee:2020vnt}. We also show that the 3point amplitude satisfies the BG equation in section \ref{null_eqs}. Finally we end with discussion and future directions in section \ref{discussions}.

\section{Notations and Conventions}

We will work in (-,+,-,+) signature in four spacetime dimensions. This is also called Klein space. The geometry of this space has been discussed in great detail in \cite{Atanasov:2021oyu}.  The scattering amplitudes, written in the boost eigenstates, behave as a correlation functions in a Lorentzian CFT on the celestial torus \cite{Atanasov:2021oyu, Atanasov:2021cje}.  We review some of the elementary equations of the celestial amplitude for this paper to be self-contained.\\

In Klein space the null momentum ($ p_i $) of $ i $-th hard massless particle is parametrized as:
\be 
p_i = \e_i \om_i q({z_i,\bar z_i})
\ee
where $\om_i$ is a real positive number, $ \e_i = + 1 (-1)$ corresponds to an outgoing (incoming) particle and
\be
q({z_i,\bar z_i}) = \{ 1+z_i \bar{z}_i, z_i+\bar{z}_i, z_i-\bar{z}_i, 1-z_i \bar{z}_i\} 
\ee
$z_i, \bar z_i$ are two real independent variables. The map from a creation/annihilation operator of a massless particle in the bulk to an operator on a celestial torus is given by the Mellin transformation
\be
\begin{gathered}
\label{mellin_phi}
\phi^{a,\e}_{h,\bar h}(z, \bar z) = \int_0^\infty d\om \, \om^{\D-1} A^a(\e \om, \s, z, \bar z) 
\end{gathered} 
\ee
where $A^a(\e \om, \s, z, \bar z)$ is an annihilaton ($ \e = + 1$)/creation ($\e=-1$) operator in the adjoint representation of an $SU(N)$ gauge theory,  $ \s $ is the helicity of the corresponding massless particle and \be h= \f{\D+\s}{2}, \, \bar h = \f{\D-\s}{2} \ee 

The momentum space amplitudes for $n$ number of external massless states written in the celestial conformal primary basis \eqref{mellin_phi} then takes the form of a 2D conformal correlator
\bea 
\no \mathcal{M}_n(\{a_i,\e_i,z_i,\bar z_i, h_i, \bar h_i\}) &=& \left< \prod_{j=1}^n \phi^{a_j,\e_j}_{h_j,\bar h_j}(z_j, \bar z_j) \right>\\
&=& \(\prod_{j=1}^n \int d\om_j \, \om_j^{\D_j-1}\) \mathcal{A}_n(a_i,\e_i\om_i, z_i,\bar z_i,\s_i)
\eea

For an on-shell massive scalar particle, the conformal primary wavefunction is obtained through the bulk-to-boundary propagator \cite{Pasterski:2016qvg,Pasterski:2017kqt}. For concreteness we take the outgoing particle with unit mass. In $ (2,2) $ signature Klein space, the momentum $ Q^\mu  $ of a massive scalar of unit mass satisfying the onshell condition $ Q^2 = -1 $ can be parametrized using the coordinates of $ \textnormal{AdS}_3/\mathbb{Z} $ as \footnote{The metric on the $ \textnormal{AdS}_3/\mathbb{Z} $ is given by
\be
ds^2_{H_3} = \f{dy^2 + dz d\bar z}{y^2} 
\ee
}
\be
Q^\mu = \f{1}{2 y}\{1+y^2 + z \bar z, z+\bar{z}, (z-\bar{z}), 1-y^2-z \bar{z} \} 
\ee

The mapping of a massive scalar particle to a conformal primary wavefunction is given through the bulk-to-boundary propagator
\be
\begin{gathered}
\Phi^\e_{\D}(X,z, \bar z) = \int_0^\infty \f{dy}{y^3} \int dz d\bar z \, G_{\D}(y,z, \bar z; w, \bar w) \, e^{i \e Q^\mu \cdot X_\mu}
\end{gathered} 
\ee
where the bulk-to-boundary propagator is given by
\be
G_{\D}(y,z, \bar z; w, \bar w) = \(\f{y}{y^2 + (z + \bar w)(z - \bar w)}\)^\D
\ee

\section{Celestial MHV Amplitudes in the Massive Scalar Background}
\label{cel_amp}

An $ n $-point MHV gluon amplitude in a massive complex scalar background was computed in \cite{Casali:2022fro}. The theory considered there was a Yang-Mills theory chirally coupled to a massive complex scalar. The massive scalar was coupled to the anti-self dual curvature tensor. An $ (n+1) $-point scalar-gluon (one scalar and $ n $ gluons) single trace, color-ordered tree-level amplitude in this set up is given by
\be 
\begin{split}
\mathcal{A}_{n+1}(\phi,1^{\e_1}_{-},2^{\e_2}_{-},3^{\e_3}_{+},\cdots ,n^{\e_n}_{+}) = \f{\left< 12 \right>^4}{\left< 12 \right>\left< 23 \right>\cdots\left< n1 \right>}\d^4\(\sum_{i=1}^n p_i + Q\)
\end{split}
\label{n+1-point-scalar-gluon}
\ee

The above amplitude is exactly the same as $ n $-point MHV amplitude except that the scalar momentum now appears in the momentum conserving delta function. The scalar field can be treated as a background and the amplitude \eqref{n+1-point-scalar-gluon} is coupled to this background by integrating over the scalar phase space \cite{Casali:2022fro}
\be
\mathcal{A}_{n}^\phi(1^{\e_1}_{-},2^{\e_2}_{-},3^{\e_3}_{+},\cdots ,n^{\e_n}_{+}) = \int \widetilde{d^3 Q} \, g(Q)\mathcal{A}_{n+1}(\phi,1^{\e_1}_{-},2^{\e_2}_{-},3^{\e_3}_{+},\cdots ,n^{\e_n}_{+}) \label{mhv_sc_bgd}
\ee
where $ g(Q) $ is the Fourier coefficient of the scalar field $ \phi(X) $ and $ \widetilde{d^3 Q} $ is the invariant measure. This amplitude is called the $ n $-point MHV amplitude in the massive scalar background and hence is denoted by $ \mathcal{A}_{n}^\phi(1^{a_1,\e_1}_{-},2^{a_2,\e_2}_{-},3^{a_3,\e_3}_{+},\cdots ,n^{a_n,\e_n}_{+}) $. We are interested in the OPE factorization of this amplitude on the celestial torus. Hence we Mellin transform the amplitude \eqref{mhv_sc_bgd} and get $ n $-point correlators on the celestial torus. To simplify calculations we will work with the 5-point scalar gluon amplitude, i.e., the 4-point MHV amplitude in the massive scalar background. \\

\section{OPE Factorisation from the 4-point Celestial Amplitude}
\label{4point}

In this section we factorize the $4$-point amplitude into $3$-point amplitude and determine the leading and subleading terms in the OPE between two positive helicity outgoing ($ \e_3=\e_4=+1 $) gluons. We show that the OPE remains the same as the MHV case \cite{Banerjee:2020vnt} . Let us start with the full 5-point scalar-gluon amplitude, given by
\bea
\no \mathcal{A}_5(\phi,1^{a_1,\e_1}_{-},2^{a_2,\e_2}_{-},3^{a_3,\e_3}_{+},4^{a_4,\e_4}_{+}) = \left\{ \mathcal{A}_4[1^{\e_1}_-2^{\e_2}_-3^{\e_3}_+4^{\e_4}_+] \ tr(T^{a_1}T^{a_2}T^{a_3}T^{a_4}) + \tn{perm (234)} \right\} \\
\times \d^{(4)}\(\sum_{i=1}^4 p_i + Q\) \hspace{1cm}
\label{5_point_mom}
\eea
where $ \mathcal{A}_4[i^{\e_i}_{\s_i}j^{\e_j}_{\s_j}k^{\e_k}_{\s_k}l^{\e_l}_{\s_l}] $ are color ordered partial MHV amplitudes given by
\be 
\mathcal{A}_4[i^{\e_i}_-j^{\e_j}_+k^{\e_k}_-l^{\e_l}_+]=\f{\left< ik \right>^4}{\left< ij \right>\left< jk \right>\left< kl \right>\left< li \right>}
\label{color_ordered}
\ee
After substituting the explicit form of the color ordered amplitude \eqref{color_ordered} in \eqref{5_point_mom} and using $ [ T^a, T^b ] = i f^{abc}T^c, \ tr(T^a T^b) = \d^{ab} $ we get
\bea
\no \mathcal{A}_5(\phi,1^{a_1,\e_1}_{-},2^{a_2,\e_2}_{-},3^{a_3,\e_3}_{+},4^{a_4,\e_4}_{+}) = \f{\left< 12 \right>^3}{\left< 23 \right>\left< 34 \right>\left< 41 \right>}\(f^{a_1 a_2 x}f^{x a_3 a_4}  - \f{z_{14}z_{23}}{z_{13}z_{24}} f^{a_1 a_3 x}f^{x a_2 a_4} \)\d^{(4)}\(\sum_{i=1}^4 p_i + Q\)\\
\label{simplified_4-point}
\eea

Substituting this expression in \eqref{mhv_sc_bgd} for the $ n=4 $ and Mellin integrating over the energies we finally get the 4-point celestial MHV amplitude in the massive scalar background, given by \cite{Casali:2022fro}
\bea
\no \widetilde{\mathcal{M}}_4^{\Phi}\(1^{a_1,\e_1}_{\D_1,-}, 2^{a_2,\e_2}_{\D_2,-}, 3^{a_3,\e_3}_{\D_3,+}, 4^{a_4,\e_4}_{\D_4,+}\) = \f{\mathcal{N}_4}{(2\pi)^4}\f{z_{12}^3}{z_{23}z_{34}z_{41}}\(f^{a_1 a_2 x}f^{x a_3 a_4}  - \f{z_{12}z_{34}}{z_{13}z_{24}} f^{a_1 a_3 x}f^{x a_2 a_4}\)\G(\D_{1}+1)\\
\no \times \, \G(\D_{2}+1)\G(\D_{3}-1)\G(\D_{4}-1) \int \widetilde{d^3 \hat{x}} (-q(z_1,\bar z_1)\cdot \hat{x})^{-\D_1-1}(-q(z_2,\bar z_2)\cdot \hat{x})^{-\D_2-1} \\
\no \times \, (-q(z_3,\bar z_3)\cdot \hat{x})^{-\D_3+1}(-q(z_4,\bar z_4)\cdot \hat{x})^{-\D_4+1} \int_0^{i\infty} d\tau \, \tau^{-1-\b_4}\phi_B(\tau)\(e^{2\pi i\b_4} - 1\)\\
\label{Mellin_4-point}
\eea
where $ \b_4 = \sum_{j=1}^4 \( {\D_j} - 1 \), \, \mathcal{N}_4 = \prod_{j=1}^4 (-i \e_j)^{\D_j-\s_j} $.
To clarify the notations let us note that a bulk point $ X^\mu $ is parameterized as $ \tau \hat x^\mu$ with $ \hat x^2 = -1 $. $ \widetilde{d^3 \hat{x}} $ is a measure on the $ \hat x^2 = -1 $ slice \footnote{For more details please see \cite{Casali:2022fro}.}. One can also compute the 3-point function in the same way and it is given by
\be 
\begin{split}
 \widetilde{\mathcal{M}}_3^{\Phi}\(1^{a_1,\e_1}_{\D_1,-}, 2^{a_2,\e_2}_{\D_2,-}, 4^{a_4,\e_4}_{\D_4,+}\) = \f{\widetilde{\mathcal{N}}_3}{(2\pi)^4}\f{2 z_{12}^3}{z_{24}z_{41}}\(i f^{a_1 a_2 a_4}\)\G(\D_{1}+1) \G(\D_{2}+1)\G(\D_{4}-1)   \\
 \times \ \int \widetilde{d^3 \hat{x}} (-q(z_1,\bar z_1)\cdot \hat{x})^{-\D_1-1} (-q(z_2,\bar z_2)\cdot \hat{x})^{-\D_2-1} (-q(z_4,\bar z_4)\cdot \hat{x})^{-\D_4+1} \\
 \times \ \int_0^{i\infty} d\tau \, \tau^{-1-\tilde{\b}_3}\phi_B(\tau)\(e^{2\pi i\tilde{\b}_3} - 1\)
\label{Mellin_3-point}
\end{split}
\ee
where \be \tilde{\b}_3 = \sum_{j=1, j\neq 3}^4 \( {\D_j} - 1 \), \, \widetilde{\mathcal{N}}_3 = \prod_{j=1, j\neq 3}^4 (-i \e_j)^{\D_j-\s_j} \label{n3}\ee.

We now take the OPE limit $ z_3 \to z_4, \, \bar z_3 \to \bar z_4 $ in \eqref{Mellin_4-point}. To do that let us first note that in the OPE limit we have
\be 
(-q(z_3,\bar z_3)\cdot \hat{x})^{-\D_3 + 1} = (-q(z_4,\bar z_4)\cdot \hat{x})^{-\D_3+1} \[ 1 + \f{\D_3 - 1}{y^2 + |z-z_4|^2}((\bar z - \bar z_4)z_{34} + (z-z_4)\bar z_{34} + z_{34}\bar z_{34}) \] + \cdots
\ee

This will be useful in the next two subsections to extract the OPE from \eqref{Mellin_4-point}. 

\subsection{Leading Order}
The leading order OPE between two positive helicity gluons was computed in \cite{Casali:2022fro} and it was shown that the leading term doesn't get any correction in the massive scalar background. Here for the sake of completeness we reproduce their results and then we move to the subleading terms in the next subsection. The leading order term of \eqref{Mellin_4-point} in the OPE expansion is 
\bea
\no \widetilde{\mathcal{M}}_4^{\Phi}\(1^{a_1,\e_1}_{\D_1,-}, 2^{a_2,\e_2}_{\D_2,-}, 3^{a_3,\e_3}_{\D_3,+}, 4^{a_4,\e_4}_{\D_4,+}\) = \f{\mathcal{N}_4}{(2\pi)^4}\f{z_{12}^3}{z_{24}z_{41}}\f{1}{z_{34}} f^{a_1 a_2 x}f^{x a_3 a_4} \G(\D_{1}+1) \G(\D_{2}+1)\G(\D_{3}-1)\G(\D_{4}-1)\\
\no \times \, \int \widetilde{d^3 \hat{x}} (-q(z_1,\bar z_1)\cdot \hat{x})^{-\D_1-1}(-q(z_2,\bar z_2)\cdot \hat{x})^{-\D_2-1} (-q(z_4,\bar z_4)\cdot \hat{x})^{-\D_3-\D_4+2}\\
 \times \, \int_0^{i\infty} d\tau \, \tau^{-1-\b_4}\phi_B(\tau)\(e^{2\pi i\b_4} - 1\)\hspace{1cm}
\label{Mellin_4-point_leading}
\eea
Replacing $ \D_4 \to \D_3 + \D_4 - 1 $ in the 3-point amplitude \eqref{Mellin_3-point}, we get
\be 
\begin{split}
 \widetilde{\mathcal{M}}_3^{\Phi}\(1^{a_1,\e_1}_{\D_1,-}, 2^{a_2,\e_2}_{\D_2,-}, 4^{a_4,\e_4}_{\D_3+\D_4-1,+}\) = \f{\mathcal{N}_4}{(2\pi)^4}\f{2 z_{12}^3}{z_{24}z_{41}}\(i f^{a_1 a_2 a_4}\)\G(\D_{1}+1) \G(\D_{2}+1) \\
 \times \ \G(\D_3+\D_{4}-2)\int \widetilde{d^3 \hat{x}} (-q(z_1,\bar z_1)\cdot \hat{x})^{-\D_1-1} (-q(z_2,\bar z_2)\cdot \hat{x})^{-\D_2-1} (-q(z_4,\bar z_4)\cdot \hat{x})^{-\D_3+\D_4+2} \\
\times \ \int_0^{i\infty} d\tau \, \tau^{-1-\b_4}\phi_B(\tau)\(e^{2\pi i\b_4} - 1\)
\label{Mellin_3-point_mod}
\end{split}
\ee

Hence at leading order we can write
\be 
\begin{gathered}
\no  \widetilde{\mathcal{M}}_4^{\Phi}\(1^{a_1,\e_1}_{\D_1,-}, 2^{a_2,\e_2}_{\D_2,-}, 3^{a_3,\e_3}_{\D_3,+}, 4^{a_4,\e_4}_{\D_4,+}\) = -\f{1}{2 z_{34}}  B(\D_3-1,\D_4-1) if^{x a_3 a_4} \widetilde{\mathcal{M}}_3^{\Phi}\(1^{a_1,\e_1}_{\D_1}, 2^{a_2,\e_2}_{\D_2}, 4^{x,\e_4}_{\D_3+\D_4-1}\)\\
\no \Rightarrow  \left< \mathcal{O}^{a_1,\e_1}_{\D_1,-}(z_1,\bar z_1) \mathcal{O}^{a_2,\e_2}_{\D_2,-}(z_2,\bar z_2)\mathcal{O}^{a_3,\e_3}_{\D_3,+}(z_3,\bar z_3)\mathcal{O}^{a_4,\e_4}_{\D_4,+}(z_4,\bar z_4)\right> \\
\no = -\f{1}{2 z_{34}} B(\D_3-1,\D_4-1) if^{x a_3 a_4} \left< \mathcal{O}^{a_1,\e_1}_{\D_1,-}(z_1,\bar z_1) \mathcal{O}^{a_2,\e_2}_{\D_2,-}(z_2,\bar z_2)\mathcal{O}^{a_4,\e_4}_{\D_3+\D_4-1,+}(z_4,\bar z_4)\right>\\
\end{gathered}
\ee

At the level of OPE the above equation reads
\be
\mathcal{O}^{a_3,+1}_{\D_3,+}(z_3,\bar z_3)\mathcal{O}^{a_4,+1}_{\D_4,+}(z_4,\bar z_4) \sim -\f{1}{2 z_{34}} B(\D_3-1,\D_4-1) if^{a_3 a_4 x}\mathcal{O}^{x,+1}_{\D_3+\D_4-1,+}(z_4,\bar z_4)
\ee


\subsection{Subleading Terms: $ \mathcal{O}(1) $}

In this section we show that the subleading ($ \mathcal{O}(1) $) term in the OPE expansion remains same as the MHV case. As we know from the study of MHV gluon amplitudes \cite{Banerjee:2020vnt}, the descendants of the leading soft gluon symmetry algebra appears at $ \mathcal{O}(1) $ in the OPE expansion. Here we only write down the action of the relevant operators on the celestial gluon amplitude. The leading conformally soft gluon operator \cite{Donnay:2018neh, Pate:2019mfs, Fan:2019emx, Nandan:2019jas, Adamo:2019ipt, Guevara:2021abz, Strominger:2021lvk, Strominger:2021mtt} for positive helicity is defined as
\be
R^{1,a}_0(z) = \lim_{\D\to 1}(\D-1)\mathcal{O}^a_{\D,+} (z,\bar z)
\ee
where $ \mathcal{O}^a_{\D,+} (z,\bar z) $ is a positive helicity primary gluon operator with scaling dimension $ \D $. The soft current  $ R^{1,a}_0(z) $ is a Kac-Moody current \cite{Strominger:2014ass, He:2014cra, Kapec:2015ena, Kapec:2014zla, Campiglia:2015qka, Nande:2017dba, He:2020ifr}. The modes of the current $ R^{1,a}_0(z) $ are denoted by $ R^{1,a}_{p,0} $. For our purpose we only mention the correlation function of the descendants $ R^{1,a}_{-p,0} \, \mathcal{O}^b_{\D,\s}(z,\bar z), \ p \geq 1 $ with a collection of gluon primaries. These are given by\footnote{Here we are using the notation of \cite{Strominger:2021lvk}.}  \cite{Banerjee:2020vnt}
\be 
\begin{split}
\left< R^{1,a}_{-p,0} \, \mathcal{O}^b_{\D,\s}(z,\bar z)\prod_{i=1}^n  \mathcal{O}^{a_i}_{\D_i,\s_i}(z_i,\bar z_i) \right> = \mathcal{R}^{1,a}_{-p,0}(z)\left< \mathcal{O}^b_{\D,\s}(z,\bar z)\prod_{i=1}^n  \mathcal{O}^{a_i}_{\D_i,\s_i}(z_i,\bar z_i) \right>
\end{split}
\ee
where the operator $ \mathcal{R}^{1,a}_{-p,0}(z) $ is defined as
\be
\begin{split}
\mathcal{R}^{1,a}_{-p,0}(z)\left< \mathcal{O}^b_{\D,\s}(z,\bar z)\prod_{i=1}^n  \mathcal{O}^{a_i}_{\D_i,\s_i}(z_i,\bar z_i) \right> = \sum_{k=1}^n \f{T^a_k}{(z_k-z)^p}\left< \mathcal{O}^b_{\D,\s}(z,\bar z)\prod_{i=1}^n  \mathcal{O}^{a_i}_{\D_i,\s_i}(z_i,\bar z_i) \right>
\label{lead_ops}
\end{split} 
\ee

Let us now consider the $ \mathcal{O}(1) $ term in the OPE expansion of \eqref{Mellin_4-point}, given by
\bea
\no \widetilde{\mathcal{M}}_4^{\Phi}\(1^{a_1,\e_1}_{\D_1,-}, 2^{a_2,\e_2}_{\D_2,-}, 3^{a_3,\e_3}_{\D_3,+}, 4^{a_4,\e_4}_{\D_4,+}\)\big|_{\mathcal{O}(1)} = \f{\mathcal{N}_4}{(2\pi)^4}\f{z_{12}^3}{z_{24}z_{41}}\G(\D_{1}+1) \G(\D_{2}+1)\G(\D_{3}-1)\G(\D_{4}-1) \\
\no \times \int \widetilde{d^3 \hat{x}} \[ \f{f^{a_1 a_2 x}f^{x a_3 a_4}}{z_{14}} + \f{f^{a_2 a_3 x}f^{x a_4 a_1}}{z_{24}} + (\D_3-1)\f{(\bar z - \bar z_4)}{y^2 + |z-z_4|^2} f^{a_1 a_2 x}f^{x a_3 a_4} \] (-q(z_1,\bar z_1)\cdot \hat{x})^{-\D_1-1} \\
\no \times \,  (-q(z_2,\bar z_2)\cdot \hat{x})^{-\D_2-1} \, (-q(z_4,\bar z_4)\cdot \hat{x})^{-\D_3-\D_4+2} \int_0^{i\infty} d\tau \, \tau^{-1-\b_4}\phi_B(\tau)\(e^{2\pi i\b_4} - 1\)\hspace{1cm}
\label{Mellin_4-point_subleading}
\eea

With the help of the operator \eqref{lead_ops}, RHS of the above equation can be written as
\be 
\begin{split}
 \widetilde{\mathcal{M}}_4^{\Phi}\(1^{a_1,\e_1}_{\D_1,-}, 2^{a_2,\e_2}_{\D_2,-}, 3^{a_3,\e_3}_{\D_3,+}, 4^{a_4,\e_4}_{\D_4,+}\)\big|_{\mathcal{O}(1)} = \f{1}{2}B(\D_3-1,\D_4-1)\[-\f{(\D_3-1)}{(\D_3+\D_4-2)}if^{x a_3 a_4}\mathcal{L}_{-1}(4)\right.\\
 \left. \times \widetilde{\mathcal{M}}_3^{\Phi}\(1^{a_1,\e_1}_{\D_1}, 2^{a_2,\e_2}_{\D_2}, 4^{x,\e_4}_{\D_3+\D_4-1}\) + \f{(\D_4-1)}{(\D_3+\D_4-2)} \mathcal{R}^{1,a_3}_{-1,0}(4) \widetilde{\mathcal{M}}_3^{\Phi}\(1^{a_1,\e_1}_{\D_1}, 2^{a_2,\e_2}_{\D_2}, 4^{a_4,\e_4}_{\D_3+\D_4-1}\) \right.\\
 \left. + \f{(\D_3-1)}{(\D_3+\D_4-2)}\mathcal{R}^{1,a_4}_{-1,0}(4)\widetilde{\mathcal{M}}_3^{\Phi}\(1^{a_1,\e_1}_{\D_1}, 2^{a_2,\e_2}_{\D_2}, 4^{a_3,\e_4}_{\D_3+\D_4-1}\)\]
\end{split}
\ee
where the argument $ (4) $ in the operators $ \mathcal{L}_{-1}, \mathcal{R}^{1,a}_{-1,0} $ implies that these modes are acting on the last particle of the 3-point amplitude. At the level of the OPE we have \footnote{We would like to emphasize that \eqref{sub} does \textit{not} hold beyond MHV sector. In the $N^kMHV$ sector the soft symmetry algebra changes because of the existence of the negative helicity soft gluons and as a result the $O(z^0\bar z^0)$ term in the OPE has to change. This is also the case for pure YM theory and has nothing to do with the existence of the massive scalar background.}
\be 
\label{sub}
\begin{split}
\mathcal{O}^{a_3,+1}_{\D_3,+}(z_3,\bar z_3)\mathcal{O}^{a_4,+1}_{\D_4,+}(z_4,\bar z_4)\big|_{\mathcal{O}(1)} \sim \f{1}{2}B(\D_3-1,\D_4-1)\[-\f{(\D_3-1)}{(\D_3+\D_4-2)}if^{x a_3 a_4} L_{-1}\right.\\
 \left. + \(\f{(\D_4-1)}{(\D_3+\D_4-2)} \d^{a_3 y}\d^{a_4 x} + \f{(\D_3-1)}{(\D_3+\D_4-2)}\d^{a_4 y}\d^{a_3 x}\) R^{1,y}_{-1,0}\] \mathcal{O}^{x,+1}_{\D_3+\D_4-1,+}(z_4,\bar z_4)
\end{split}
\ee
One can recognize that this is the $ \mathcal{O}(1) $ term in the OPE obtained by \cite{Banerjee:2020vnt,Ebert:2020nqf} in the MHV case. Thus we can see that $ \mathcal{O}(1) $ OPE also doesn't change in the presence of the massive scalar background.

\section{Subleading Soft Gluon Theorem in Massive Scalar  Background}
\label{sub_th}

In this section we show that the subleading conformal soft gluon theorem remains same in a massive scalar background. More precisely, we show that the subleading conformal soft limit, $\D_{4}\rightarrow 0$, of \eqref{Mellin_4-point} is equivalent to the action of the subleading soft operator \cite{Donnay:2018neh, Pate:2019mfs, Fan:2019emx, Nandan:2019jas, Adamo:2019ipt} of the $4$-th particle, on the $3$-pt correlation function \eqref{Mellin_3-point}.
Hence we start with the 4-point amplitude \eqref{Mellin_4-point} and take the conformal soft limit $\D_{4}\rightarrow 0$ to get
\bea
\no \lim_{\Delta_{4}\rightarrow 0}\ \Delta_{4} \widetilde{\mathcal{M}}_4^{\Phi}\(1^{a_1,\e_1}_{\D_1,-}, 2^{a_2,\e_2}_{\D_2,-}, 3^{a_3,\e_3}_{\D_3,+}, 4^{a_4,\e_4}_{\D_4,+}\) = -i \f{\mathcal{N}_3}{(2\pi)^4}\f{z_{12}^3}{z_{23}z_{34}z_{41}}\(f^{a_1 a_2 x}f^{x a_3 a_4}  - \f{z_{12}z_{34}}{z_{13}z_{24}} f^{a_1 a_3 x}f^{x a_2 a_4}\)\\
\no \times \, \G(\D_{1}+1)\G(\D_{2}+1)\G(\D_{3}-1) \int \widetilde{d^3 \hat{x}} (-q(z_1,\bar z_1)\cdot \hat{x})^{-\D_1-1}(-q(z_2,\bar z_2)\cdot \hat{x})^{-\D_2-1} \\
\no \times \, (-q(z_3,\bar z_3)\cdot \hat{x})^{-\D_3+1}(-q(z_4,\bar z_4)\cdot \hat{x}) \int_0^{i\infty} d\tau \, \tau^{-\b_3}\phi_B(\tau)\(e^{2\pi i\b_3} - 1\)\\
\label{Mellin_4-point_1}
\eea
The above result can be written as
\begin{equation}\label{eq:38}
\begin{split}
\lim_{\Delta_{4}\rightarrow 0}\ \Delta_{4}\mfour
 = \bigg(\frac{c_{1}}{z_{14}} + \frac{c_{2}}{z_{24}} + \frac{c_{3}}{z_{34}}\bigg)\ i \frac{{\mathcal{N}_{3}}}{(2\pi)^{4}}\ \frac{z_{12}^{3}}{z_{23}z_{31}} \\
 \times \, \Gamma(\Delta_{1}+1)\Gamma(\Delta_{2}+1)\Gamma(\Delta_{3}-1) \int \widetilde{d^3 \hat{x}} (-q(z_1,\bar z_1)\cdot \hat{x})^{-\D_1-1}(-q(z_2,\bar z_2)\cdot \hat{x})^{-\D_2-1} \\
 \times \, (-q(z_3,\bar z_3)\cdot \hat{x})^{-\D_3+1}(-q(z_4,\bar z_4)\cdot \hat{x}) \int_0^{i\infty} d\tau \, \tau^{-\b_3}\phi_B(\tau)\(e^{2\pi i\b_3} - 1\)
\end{split}
\end{equation}
where $c_{1} = f^{x a_{2}a_{3}}f^{a_4 a_{1} x}$, $c_{2} = f^{a_{1} x a_{3}}f^{a_4 a_{2} x }$ and $c_{3} = f^{a_{1}a_{2}x}f^{a_{4}a_{3} x}$.\\

We shall now show that the above expression for the conformal soft limit of the celestial correlation function is the same as the action of the subleading soft operator on the $3$-pt correlation function. The subleading soft gluon theorem in Mellin space is given by \cite{Pate:2019lpp},
\begin{equation}
\begin{split}
\left< R^{0,a_{4}}(z_4,\bar z_4) \prod_{i=1}^3 \mathcal{O}_{\D_i, \s_i}^{a_i,\e_i}(z_i,\bar z_i)\right> = -\sum_{k=1}^{3}\ \frac{\e_k}{z_{4k}}(-2\bar{h}_{k} + 1 +\bz_{4k}\bar{\p}_{k})\ T^{a_{4}}_k \mathcal{P}^{-1}_{k}\left< \prod_{i=1}^3\mathcal{O}_{\D_i, \s_i}^{a_i,\e_i}(z_i,\bar z_i)\right>
\label{sub_lead_soft}
\end{split}
\end{equation}
where $R^{0,a}(z,\bar z)$ is the subleading conformally soft gluon operator defined by \be R^{0,a}(z,\bar z) = \lim_{\D \to 0}\D \mathcal{O}^a_{\D,+}(z,\bar z) \ee
  $T^{a}_k $ is the lie algebra generator in the adjoint representation of the gauge group and $\mathcal{P}^{-1}_{k}$ is a dimension lowering operator acting on the $k$-th primary field. The action of both is 
\begin{equation}
T^{a}_{k}\mathcal{O}^{a_{i},\e_i}_{\D_{i},\s_i}(z_i,\bar z_i) = if^{a a_{i} x}\, \mathcal{O}^{x,\e_i}_{\D_{i},\s_i}(z_i,\bar z_i)\, \d_{k,i} ,\ \mathcal{P}^{-1}_{k}\mathcal{O}^{a_{i},\e_i}_{\D_{i},\s_i}(z_i,\bar z_i) = \mathcal{O}^{a_i,\e_i}_{\D_{i}-1,\s_i}(z_i,\bar z_i)\d_{ki}
\end{equation}
 By repeated use of the following equation,
\begin{equation}
(-q_{i}\cdot \hat{x}) = \frac{y^{2} + (z-z_{i})(\bz - \bz_{i})}{y}
\end{equation}
and the explicit expression for 3-point function given in \eqref{Mellin_3-point} on the RHS of \eqref{sub_lead_soft}, one can show that (appendix \ref{sublead_gluon})
\be
\begin{split}
-\sum_{k=1}^{3}\ \frac{\e_k}{z_{4k}}(-2\bar{h}_{k} + 1 +\bz_{4k}\bar{\p}_{k})\ T^{a_{4}}_k \mathcal{P}^{-1}_{k}\left< \prod_{i=1}^3\mathcal{O}_{\D_i, \s_i}^{a_i,\e_i}(z_i,\bar z_i)\right> = \bigg(\frac{c_{1}}{z_{14}} + \frac{c_{2}}{z_{24}} + \frac{c_{3}}{z_{34}}\bigg)\ i \frac{{\mathcal{N}_{3}}}{(2\pi)^{4}}\ \frac{2 z_{12}^{3}}{z_{23}z_{31}} \\
 \times \, \Gamma(\Delta_{1}+1)\Gamma(\Delta_{2}+1)\Gamma(\Delta_{3}-1) \int \widetilde{d^3 \hat{x}} (-q(z_1,\bar z_1)\cdot \hat{x})^{-\D_1-1}(-q(z_2,\bar z_2)\cdot \hat{x})^{-\D_2-1} \\
 \times \, (-q(z_3,\bar z_3)\cdot \hat{x})^{-\D_3+1}(-q(z_4,\bar z_4)\cdot \hat{x}) \int_0^{i\infty} d\tau \, \tau^{-\b_3}\phi_B(\tau)\(e^{2\pi i\b_3} - 1\)
 \label{sublead_rhs}
\end{split} 
\ee

Comparing \eqref{eq:38}, \eqref{sub_lead_soft} and \eqref{sublead_rhs} we get
\begin{equation}
\begin{split}
\lim_{\Delta_{4}\rightarrow 0}\ \Delta_{4}\mfour = \left< R^{0,a_{4}}(z_4, \bar z_4) \prod_{i=1}^3 \mathcal{O}_{\D_i, \s_i}^{a_i,\e_i}(z_i,\bar z_i)\right> \\
= -\f{1}{2}\sum_{k=1}^{3}\ \frac{\e_k}{z_{4k}}(-2\bar{h}_{k} + 1 +\bz_{4k}\bar{\p}_{k})\ T^{a_{4}}_k \mathcal{P}^{-1}_{k}\left< \prod_{i=1}^3\mathcal{O}_{\D_i, \s_i}^{a_i,\e_i}(z_i,\bar z_i)\right>
\end{split}
\end{equation}

Thus we see that subleading soft gluon theorem for a positive helicity soft gluon does not change if we couple the Yang-Mills with the massive complex scalar in a chiral way mentioned in section \ref{cel_amp}.\\

\subsection{BG Equations in Massive Scalar Background}
\label{bgeqn}
A set of differential equations for the gluon MHV amplitudes were obtained in \cite{Banerjee:2020vnt} by demanding the consistency between the subleading soft gluon theorem and OPE factorisation  at  $\mathcal{O}(1)$ \footnote{Although these equations look very similar to the KZ equation which appears in WZW models, they are qualitatively different. In particular, they cannot be derived by any Sugawara construction on the celestial sphere. One way to see this is the following. Sugawara construction leads to the quantization of dimensions of primary operators which we know is not the case in celestial holography.}. Here we have shown that, even in the case of MHV amplitudes in a massive scalar background the subleading soft gluon theorem and the OPE factorisation at $\mathcal{O}(1)$ do not change. Hence we conclude that the celestial MHV amplitudes coupled to a massive scalar background also satisfy BG equations. This is not surprising because the existence of the null state, which gives rise to the differential equations, is guaranteed by the leading and the subleading soft gluon theorems and $ SL(2,C) $ invariance. The presence of a massive dilaton background breaks the scaling as well as translational invariance but the soft gluon theorems remain unchanged.  So the BG equations should also not change. 


\section{Discussions}
\label{discussions}

In this paper we extracted the OPE between two positive helicity outgoing gluons from the Mellin amplitude of the Yang-Mills theory chirally coupled to massive scalar background. The leading order term was already computed in \cite{Casali:2022fro}. Here we have computed a subleading term. One of our motivation behind this work was to check if the scattering amplitudes in this theory is also a solution of the BG equations. We have shown in this paper that this is indeed the case\footnote{This seems to raise a puzzle because we are saying that the MHV amplitudes in pure YM and the YM coupled to massive dilaton both satisfy the BG equation. The resolution of this is the following: The MHV amplitudes in pure YM are not only the solutions of BG equations but they also have to satisfy the Ward identities coming from the space time translation invariance and the scale invariance of the YM theory. But when coupled to the massive dilaton background both these Ward identities are no longer valid and so we get a different kind of MHV amplitude by solving the BG equation.}. Though the scaling and translation symmetry were explicitly broken for the theory we considered, the leading and subleading soft theorems remain unchanged. The OPE factorization at $ \mathcal{O}(1) $ is completely determined in terms of the descendants of the $ SL(2,C) $ and the leading soft symmetry algebra in the same way as the MHV amplitudes. On the other hand we have also shown that the subleading soft gluon theorem does not change in the massive background. Thus by comparing the $ \mathcal{O}(1) $ OPE with the subleading soft gluon theorem we get the same BG equations as the MHV amplitudes. More generally, we can say that the scattering amplitudes of all the theories which respect the symmetries coming from the leading and subleading soft gluon theorems should satisfy the BG equations.

However, if one considers the graviton scattering amplitude and breaks some of the symmetries considered here, the situation will change completely. This is because of the fact that the leading soft graviton theorem is a consequence of supertranslation invariance. So if in a gravitational theory the translation symmetry is broken, the leading soft graviton theorem would no longer holds. Then one can ask the question that what happens to the decoupling equations obtained for the MHV graviton scattering amplitudes in \cite{Banerjee:2020zlg}. 

It would also be interesting to check how the analysis of our work would change in the context of deformed soft algebras for gauge theories recently considered in \cite{Melton:2022fsf}. We hope to answer some of these questions in the near future.

\section{Acknowledgement}
The work of SB is partially supported by the  Swarnajayanti Fellowship (File No- SB/SJF/2021-22/14) of the Department of Science and Technology and SERB, India and by SERB grant MTR/2019/000937 (Soft-Theorems, S-matrix and Flat-Space Holography). PP would like to thank the organizers and the participants of the CHEP in-house symposium, 2022 held in IISc, Bangalore where part of this work was presented. The work of PP is supported by an IOE endowed postdoctoral position at IISc, Bangalore, India.

\appendix

\section{Subleading Soft Gluon Theorem}
\label{sublead_gluon}

In this appendix we derive \eqref{sublead_rhs}.  Let us start with the $3$-point correlation function given by \eqref{Mellin_3-point},
\be 
\begin{split}
 \widetilde{\mathcal{M}}_3^{\Phi}\(1^{a_1,\e_1}_{\D_1,-}, 2^{a_2,\e_2}_{\D_2,-}, 3^{a_3,\e_3}_{\D_3,+}\) = \f{{\mathcal{N}}_3}{(2\pi)^4}\f{2 z_{12}^3}{z_{23}z_{31}}\(i f^{a_1 a_2 a_3}\)\G(\D_{1}+1) \G(\D_{2}+1)\G(\D_{3}-1)   \\
 \times \ \int \widetilde{d^3 \hat{x}} (-q(z_1,\bar z_1)\cdot \hat{x})^{-\D_1-1} (-q(z_2,\bar z_2)\cdot \hat{x})^{-\D_2-1} (-q(z_3,\bar z_3)\cdot \hat{x})^{-\D_3+1} \\
 \times \ \int_0^{i\infty} d\tau \, \tau^{-1-\b_3}\phi_B(\tau)\(e^{2\pi i\b_3} - 1\)
 \label{3pt11}
\end{split}
\ee
where \be \b_3 = \sum_{j=1}^3 \( {\D_j} - 1 \), \, {\mathcal{N}}_3 = \prod_{j=1}^3 (-i \e_j)^{\D_j-\s_j} \label{nt3}\ee.

So, the action of the subleading soft operator on the $3$-point correlation function is
\begin{equation}
\begin{split}\label{eq:313}
& -\sum_{k=1}^{3}\ \frac{\e_k}{z_{4k}}(-2\bar{h}_{k} + 1 +\bz_{4k}\bar{\p}_{k})\ T^{a_{4}}_k \mathcal{P}^{-1}_{k} \widetilde{\mathcal{M}}_3^{\Phi}\(1^{a_1,\e_1}_{\D_1,-}, 2^{a_2,\e_2}_{\D_2,-}, 3^{a_3,\e_3}_{\D_3,+}\)\\ 
& = \e_1 \frac{c_{1}}{z_{41}}(-\Delta_{1} +\bz_{41}\bar{\p}_{1})\ \widetilde{\mathcal{M}}_{3}^{\Phi}(1_{\Delta_{1}-1,-},2_{\Delta_{2},-},3_{\Delta_{3},+}) +\e_2 \frac{c_{2}}{z_{42}}(-\Delta_{2} +\bz_{42}\bar{\p}_{2})\ \widetilde{\mathcal{M}}_{3}^{\Phi}(1_{\Delta_{1},-},2_{\Delta_{2}-1,-},3_{\Delta_{3},+})\\
& \hspace{3cm}+\e_3 \frac{c_{3}}{z_{43}}(-\Delta_{3} + 2 +\bz_{43}\bar{\p}_{3})\ \widetilde{\mathcal{M}}_{3}^{\Phi}(1_{\Delta_{1},-},2_{\Delta_{2},-},3_{\Delta_{3}-1,+})
\end{split}
\end{equation}
We have pulled out the colour factors ($i f$) out of the amplitude and $\widetilde{\mathcal{M}}^{\Phi}_{3}(1_{\Delta_{1}-1,-},2_{\Delta_{2},-},3_{\Delta_{3},+}) $ etc. are the colour stripped amplitude in eq.\eqref{eq:313}. The right hand side of eq.\eqref{eq:313} can be written as
\begin{equation}
\begin{split}
 \bigg(\frac{c_{1}}{z_{14}} + \frac{c_{2}}{z_{24}} + \frac{c_{3}}{z_{34}} + (c_1+c_2+c_3)\pa_4 \bigg)\ i\f{{\mathcal{N}}_3}{(2\pi)^4}\f{2 z_{12}^3}{z_{23}z_{31}} \G(\D_{1}+1) \G(\D_{2}+1)\G(\D_{3}-1)   \\
 \times \ \int \widetilde{d^3 \hat{x}} (-q(z_1,\bar z_1)\cdot \hat{x})^{-\D_1-1} (-q(z_2,\bar z_2)\cdot \hat{x})^{-\D_2-1} (-q(z_3,\bar z_3)\cdot \hat{x})^{-\D_3+1} (-q(z_4,\bar z_4)\cdot \hat{x})\\
 \times \ \int_0^{i\infty} d\tau \, \tau^{-1-\b_3}\phi_B(\tau)\(e^{2\pi i\b_3} - 1\)
\end{split}
\end{equation}

where we have used the following equation
\be
\bar z_{4k} \bar \pa_k(-q_k \cdot \hat x) = - (-q_k \cdot \hat x)+(1+z_{k4}\pa_4)(-q_4 \cdot \hat x)
\ee

Now using the Jacobi identity \be c_1+c_2+c_3 = 0 \ee we finally get 
\begin{equation}
\begin{split}
-\sum_{k=1}^{3}\ \frac{\e_k}{z_{4k}}(-2\bar{h}_{k} + 1 +\bz_{4k}\bar{\p}_{k})\ T^{a_{4}}_k \mathcal{P}^{-1}_{k} \widetilde{\mathcal{M}}_3^{\Phi}\(1^{a_1,\e_1}_{\D_1,-}, 2^{a_2,\e_2}_{\D_2,-}, 3^{a_3,\e_3}_{\D_3,+}\) = \bigg(\frac{c_{1}}{z_{14}} + \frac{c_{2}}{z_{24}} + \frac{c_{3}}{z_{34}} \bigg)\ i\f{{\mathcal{N}}_3}{(2\pi)^4}\f{2 z_{12}^3}{z_{23}z_{31}} \\
\times \G(\D_{1}+1) \G(\D_{2}+1)\G(\D_{3}-1)   \int \widetilde{d^3 \hat{x}} (-q(z_1,\bar z_1)\cdot \hat{x})^{-\D_1-1} (-q(z_2,\bar z_2)\cdot \hat{x})^{-\D_2-1} \ (-q(z_3,\bar z_3)\cdot \hat{x})^{-\D_3+1} \\
 \times (-q(z_4,\bar z_4)\cdot \hat{x}) \ \int_0^{i\infty} d\tau \, \tau^{-\b_3}\phi_B(\tau)\(e^{2\pi i\b_3} - 1\)
\end{split}
\end{equation}
which is same as \eqref{sublead_rhs}.

\section{Solution of the BG Equations}
\label{null_eqs}
In subsection \ref{bgeqn}, we have argued that the BG equations remain same if we put the MHV amplitudes in a massive scalar background. Here we show explicitly that the three point MHV amplitude in the massive background satisfies the BG equations. We first write down the most general form of the 3-point amplitude using the $ SL(2,C) $ symmetry and then derive the constraints for the 3-point coefficient imposed by the BG equations.  \\

Let us start with the color ordered $ SL(2,C) $-covariant 3-point amplitude given by,
\bea
\no \widetilde{\mathcal{M}}_3(1^-_{\D_1} 2^-_{\D_2} 3^+_{\D_3}) = C(\D_1,\D_2,\D_3) z_{12}^{h_3-h_1-h_2}z_{13}^{h_2-h_1-h_3}z_{23}^{h_1-h_2-h_3} \bar z_{12}^{\bar h_3 - \bar h_1-\bar h_2}\bar z_{13}^{\bar h_2-\bar h_1-\bar h_3}\bar z_{23}^{\bar h_1-\bar h_2-\bar h_3}\\
\label{m_g_3}
\eea


There are two sets of decoupling equations for the color ordered amplitudes \cite{Banerjee:2020vnt,Hu:2021lrx}. They are
\be 
\begin{split}
\(\pa_3 -\f{\D_3}{z_{13}} - \f{1}{z_{23}}\)\widetilde{\mathcal{M}}_3(1^-_{\D_1} 2^-_{\D_2} 3^+_{\D_3})+\e_1\e_3 \f{\D_1-\s_1-1+\bar z_{13}\bar \pa_1}{z_{13}}\widetilde{\mathcal{M}}_3(1^-_{\D_1-1} 2^-_{\D_2} 3^+_{\D_3+1})=0\\
\(\pa_3 -\f{\D_3}{z_{23}} - \f{1}{z_{13}}\)\widetilde{\mathcal{M}}_3(1^-_{\D_1} 2^-_{\D_2} 3^+_{\D_3})+\e_2\e_3 \f{\D_2-\s_2-1+\bar z_{23}\bar \pa_2}{z_{23}}\widetilde{\mathcal{M}}_3(1^-_{\D_1} 2^-_{\D_2-1} 3^+_{\D_3+1})=0
\end{split}
\label{null_eqn}
\ee


Using \eqref{m_g_3} in \eqref{null_eqn} we get the following constraints on the 3-point coefficient
\bea
\label{rec_rel1}
C(\D_1-1,\D_2,\D_3+1) = \e_1\e_3 \f{(\D_1-\D_2-\D_3+1)}{(\D_3-\D_1-\D_2-1)}C(\D_1,\D_2,\D_3)\\
C(\D_1,\D_2-1,\D_3+1) = \e_2\e_3 \f{(\D_2-\D_1-\D_3+1)}{(\D_3-\D_1-\D_2-1)}C(\D_1,\D_2,\D_3)
\label{rec_rel2}
\eea



Now, one can check that the 3-point coefficient given by \cite{Casali:2022fro}
\be
 C(\D_1,\D_2,\D_3) = \mathcal{N}_3 \, \G\(\f{\D_1+\D_2-\D_3+3}{2}\)\G\(\f{\D_1-\D_2+\D_3-1}{2}\)\G\(\f{-\D_1+\D_2+\D_3-1}{2}\) f(\b)
 \label{strominger}
\ee

satisfies \eqref{rec_rel1}, \eqref{rec_rel2}, where $ f(\b) $ is any function with $ \b = \sum_{i=1}^3 \D_i $ and $ \mathcal{N}_3 $ is given by \eqref{nt3}.

\end{document}